%% file: main.tex
\documentclass[a4paper,11pt]{article}
\pdfoutput=1 
\usepackage{aas_macros}
\usepackage{jcappub} 
\usepackage[T1]{fontenc} 

\newcommand{\hMpc}{\mathrm{h}^{-1}\mathrm{Mpc}}

\usepackage{orcidlink}
\usepackage{multirow}
\usepackage{amsmath}
\usepackage{bbm}
\usepackage{stmaryrd}
\usepackage{bm}
\usepackage{graphicx}  
\usepackage{tikz}      
\usetikzlibrary{positioning, decorations.pathreplacing} 
\usepackage{subcaption}
\usepackage{adjustbox}
\usepackage{comment}

\usepackage{color,hyperref}

\title{\boldmath Cosmic Shear constraints from HSC Year 3 with clustering calibration of the tomographic redshift distributions from DESI}

\author[1,2]{J. Choppin de Janvry\,\orcidlink{0009-0008-8066-446X}}
\author[3]{B. Dai,}
\author[1,4]{S. Gontcho A Gontcho,}
\author[1,2]{U. Seljak,}
\author[5]{T. Zhang\,\orcidlink{0000-0002-5596-198X}}

\affiliation[1]{Physics Division, Lawrence Berkeley National Laboratory, Berkeley, CA 94720, USA}
\affiliation[2]{Department of Physics, University of California, Berkeley, CA 94720, USA}
\affiliation[3]{School of Natural Sciences, Institute for Advanced Study, 1 Einstein Drive, Princeton, New
Jersey 08540, USA}
\affiliation[4]{Department of Astronomy, University of Virginia, Charlottesville, VA 22904, USA}
\affiliation[5]{Department of Physics and Astronomy and PITT PACC, University of Pittsburgh, Pittsburgh, PA 15260, USA}

\emailAdd{jean.choppindejanvrydev@gmail.com}
\emailAdd{biwei@ias.edu}
\emailAdd{satya@virginia.edu}
\emailAdd{useljak@berkeley.edu}
\emailAdd{tq.zhang@pitt.edu}

\abstract{We reanalyze cosmological constraints from Hyper Suprime-Cam (HSC) Y3 
shear-shear correlation function using new calibration of the tomographic redshift distribution via the clustering redshifts method with DESI spectroscopy presented in Choppin de Janvry {\it et al.} (2025a) \cite{ChoppinDeJanvry2025a}.  
We present both importance sampling of the original MCMC chains by HSC, applying the weights of our newly calibrated $\Delta z$ priors, as well as full MCMC analysis with new photometric redshift distributions, finding consistent results between the two. 
We obtain the growth of structure parameter $S_8\equiv\sigma_8\sqrt{\Omega_m/0.3}=0.804^{+0.019}_{-0.018}$, compared to previous HSC Y3 result of $S_8=0.769^{+0.031}_{-0.034}$, which is a 1.8 reduction of error due to the improved clustering redshift calibrations, with the central value shifting considerably higher towards Planck cosmology. 
With the new photometric redshift calibration, HSC Y3 has comparable constraining power to the recent KIDS Legacy and DES Y6 results.}

\begin{document} 
\maketitle
\flushbottom

\input{sections/introduction}
\input{sections/data}

\input{sections/method}

\input{sections/results}

\input{sections/conclusion}

\input{sections/acknowledgements}

\newpage
\appendix
\input{sections/appendix}

\newpage

\bibliographystyle{JHEP}
\bibliography{src/intro,src/desi,src/hsc,src/method,src/software,src/references,src/s8,src/others}
 
\end{document}

%% file: sections/introduction.tex
\section{Introduction}
\label{sec:intro}
Weak gravitational lensing (WL) causes subtle distortion of observed galaxy shapes due to the gravitational influence of intervening large-scale structures along the line of sight. By statistically analyzing correlations in these shear distortions across large samples of galaxies, weak lensing surveys provide a powerful probe of the large-scale structure of the universe \citep{Bartelmann_2001, Kilbinger_2015}. These analyses allow us to constrain key cosmological parameters, particularly $\Omega_\mathrm{m}$ (matter density) and $\sigma_8$ (amplitude of matter fluctuations), test the standard $\Lambda$CDM model and explore potential deviations from it. WL causes distortions of cosmic microwave background, with the latest constraints presented in \cite{Qu2025_ACTSPTPLANCK_CMBL_S8}. WL cross-correlation with galaxies can be combined with galaxy auto-correlation for additional  constraints on cosmological parameters. This technique however suffers from potential systematics in the modeling of the cross-correlation coefficient between galaxies and dark matter \cite{Guzik01}, as well as from observational systematics in the galaxy auto-correlation. Therefore, in this paper we focus on WL analyses without galaxy clustering. 

Several Stage-III weak lensing surveys, including the Hyper Suprime-Cam (HSC; \cite{HSC_overview, HSC2022datarelease3, HSCShapeCataloguePDR3Li2022}), the Kilo-Degree Survey (KiDS; \cite{KiDS2025datarelease5, 2012KiDS_overview}), and the Dark Energy Survey (DES; \cite{DES_overview, desy6_gcwl}), have made significant progress in constraining the cosmological model. However, these weak lensing surveys have consistently reported values for the lensing amplitude $S_8\equiv\sigma_8\sqrt{\Omega_\mathrm{m}/0.3}$ that are systematically lower than those inferred from cosmic microwave background (CMB) measurements \cite{Planck2018_CMB, Louis2025_ACTDR6_PS_S8} under the $\Lambda$CDM framework \cite{DESY3Amon,DESY3Secco,DESy6_S8_CosmicShear, Li2023_HSCY3_CosmicShear, Asgari2021_KiDS-1000_CosmicShear, DES_KiDS}. This discrepancy, often described as the $S_8$ tension, has led to substantial interest, as it may point to either unknown systematics in the analyses \citep{Leonard2024photometric,Amon2022nonlinear} or new physics beyond the standard model \citep{Abellan2021linear}. 

Disentangling these possibilities requires improved control of systematics, particularly in the context of next-generation weak lensing surveys, such as the Rubin Observatory Legacy Survey of Space and Time (LSST) \citep{LSST2019survey}, the Nancy Grace Roman Space Telescope \citep{roman2019wfirst}, and Euclid \citep{EuclidOverview}. These surveys will achieve unprecedented precision, making the mitigation of systematics a crucial priority. Among various systematics, photometric redshift calibration is particularly critical, as errors in redshift distributions propagate directly into biases in the constraints of $S_8$.

Most recently, the KiDS survey introduced several enhancements to its analysis in the final KiDS-Legacy release, including improved photometric redshift calibration, improved image reduction, and updated intrinsic alignment modeling \citep{Wright2025_KiDSLegacy, wright2025kidslegacyClusterZ, KiDS2025datarelease5}. Among these, improvements in photometric redshift calibration played a particularly significant role, leading to tighter constraints on cosmological parameters and a notable shift in their $S_8$ results. The Hyper Suprime-Cam (HSC) Year 3 (Y3) analysis, on the other hand, adopted a wide uniform prior on photometric redshift biases due to limited photo-z calibration in higher redshift bins \citep{Li2023_HSCY3_CosmicShear, Dalal2023CosmoShearHSC}. This approach allowed HSC to account for uncertainties in the redshift distributions and enabled robust inference, but it also resulted in weaker constraints on $S_8$. Both KiDS and HSC highlight the critical role of photometric redshift calibration in reducing systematic uncertainties and improving the precision of weak lensing cosmology.

In this work, we reanalyze the HSC Y3 shear-shear correlation function using the updated clustering redshift calibration presented in \cite{ChoppinDeJanvry2025a}, which leverages spectroscopic data from the Dark Energy Spectroscopic Instrument (DESI, \cite{DESI2016b.Instr, DESI2022.KP1.Instr}). The aforementioned analysis has obtained tighter priors on the mean redshift shift for all four tomographic bins of the cosmic shear analyses from HSC Y3. The improved redshift calibration is compared to the original analysis \cite{HSCClusteringRau2023} as well as other constraints obtained on the mean shift of the two last redshift bins. For example, the results obtained show good consistency with lensing shear ratio calibration (SR) \cite{Rana2025HSCY3CosmicShearRatios}, as presented in the companion paper \cite{ChoppinDeJanvry2025a}. SR \cite{Rana2025HSCY3CosmicShearRatios, Emas_2024_shear_ratio,Schneider_2016_shear_ratio} provide alternative information to calibrate the photo-$z$ distributions of HSC, by probing the tangential shear around spectroscopic lenses. These shear ratios are sensitive to the positions of the source galaxies, hence probing the mean redshifts of the tomographic bins. The mean shifts ($\Delta z_i$ where $i$ is the index of the tomographic bin) are also compared to the shifts obtained using the "self-calibration" approach used in \cite{Dalal2023CosmoShearHSC, Li2023_HSCY3_CosmicShear}. This method places a wide uniform prior on the $\Delta z_i$ parameters that are not sufficiently well constrained. The re-analysis found that the central value can lie on the $\sim1-2\sigma$ edge of the posterior distributions found by the self-calibration approach, thus motivating a re-analysis of the cosmic shear constraints.

Therefore, the goal of this work is to quantify the impact of this improved redshift calibration on the cosmological constraints derived from HSC Y3 and to explore its implications for the $S_8$ tension. 

This paper is organized as follows. In Section \ref{sec:data}, we describe the dataset used in this paper: the HSC Y3 shape catalog and the galaxy redshift distribution $n(z)$ calibrated with the DESI data. In Section \ref{sec:method}, we present the methodology followed in this analysis to recover cosmological constraints. Section \ref{sec:results} showcases the cosmological constraints inferred. Finally, we conclude this work in Section \ref{sec:conclusion}. We present systematic checks and additional information in appendices \ref{sec:appendix:params}, \ref{sec:appendix:chain_processing}, \ref{sec:appendix:combined}.

%% file: sections/data.tex
\section{Datasets}
\label{sec:data}

This section describes the datasets and catalogs used in this analysis. We first present the HSC Y3 shape catalog in section \ref{sec:data:hsc}, then present the updated $n(z)$ distributions derived with DESI spectroscopic data in section \ref{sec:data:desi}.



\subsection{HSC Y3 Shape Catalog}
\label{sec:data:hsc}

The Hyper Suprime-Cam Subaru Strategic Program (HSC-SSP) is a wide and deep imaging survey conducted by the 8.2-meter Subaru Telescope. The weak lensing shear catalog used in this study is the HSC S19A galaxy shape catalog, i.e., the so-called HSC Year 3 shear catalog\cite{HSCShapeCataloguePDR3Li2022}, an intermediate release between the HSC Public Data Release 2 (PDR2) \cite{hsc_pdr2} and HSC Public Data Release 3 (PDR3) \cite{HSC2022datarelease3}. This catalog builds upon $i_{\mathrm{band}}$ imaging data obtained from 2014 to 2019 in the Wide fields of the HSC survey, at a mean $i_{\mathrm{band}}$ PSF Full-Width of Half-Maximum (FWHM) of $0.59''$ \cite{HSCShapeCataloguePDR3Li2022}. A conservative $i_{\mathrm{band}}<24.5$ magnitude cut is applied to the dataset. Even with the conservative magnitude cut, the catalog has an effective number density of $19.9 {\;\rm arcmin}^{-2}$. The geometric footprint of the catalog is split into 6 fields: \texttt{HECTOMAP}, \texttt{XMM}, \texttt{VVDS}, \texttt{GAMA09H}, \texttt{GAMA15H} and \texttt{WIDE12H}.

We adopt consistent tomographic binning to the HSC Y3 cosmic shear \cite{Li2023_HSCY3_CosmicShear, Dalal2023CosmoShearHSC}, tomographic $2\times2$pt\cite{Zhang2025aTomographic2x2pt} and $3\times2$pt analysis\cite{Zhang2025bTomographic3x2pt}. The tomographic binning is defined using the photometric redshifts inferred by \texttt{DNNz} (Nishizawa \textit{et al}, in preparation). \texttt{DNNz} is a neural network to estimate conditional densities for photometric redshifts, with 100 output nodes (each representing a redshift bin) and 5 hidden layers using multilayer perceptrons \cite{HSCClusteringRau2023}.

Specifically, the bin edges ($z_\mathrm{p}=0.3, 0.6, 0.9, 1.2,1.5$) use $z_{\mathrm{best}}^{\texttt{DNNz}}$ to assign galaxies into four tomographic bins. The $_\mathrm{best}$ subscript notes the redshift point estimate that minimizes the risk function, $R(z_{\mathrm{p}})=\int P(z)L(\frac{z_{\mathrm{p}}-z}{1+z})\mathrm{d}z$, based on the individual probability density function $P(z)$ obtained through $\texttt{DNNz}$, and with $L(x)$ a loss function \cite{HSCPHotozTanaka2017, PhotoZHSCNishizawa}.
In addition, a \textit{quality control cut} is applied to the first and second tomographic bins, which depends on the $95\%$ confidence interval boundaries of the \texttt{DNNz} and \texttt{Mizuki} \cite{PhotoZHSCNishizawa, MizukiHSCTanaka2015, HSCPHotozTanaka2017} photometric redshift algorithms,
\begin{equation}\label{eq:calib_cut}
(z^{\texttt{DNNz}}_{95,\,\mathrm{max}}-z^{\texttt{DNNz}}_{95,\,\mathrm{min}})<2.7\;\;\&\;\;(z^{\texttt{Mizuki}}_{95,\,\mathrm{max}}-z^{\texttt{Mizuki}}_{95,\,\mathrm{min}})<2.7.
\end{equation}
This selection is performed in order to remove possible galaxies with photo-$z$ probability density functions reporting multi-modal distributions with a second, smaller peak around $z\sim3$. The cut removes $\sim31\%$ of sources in the first bin, and $\sim8\%$ of sources in the second bin \cite{HSCClusteringRau2023}. \texttt{Mizuki} \cite{MizukiHSCTanaka2015} is a photometric redshift code using spectral fitting of templates to the galaxy photometry.

In the rest of this analysis, when using redshifts outside of the HSC tomographic bins, the quality control cut is also applied to all sources with redshifts between $z_\mathrm{p}\sim0$ and $z_\mathrm{p}\sim0.3$, which removes about $\sim16\%$ of the sources within that redshift range. Therefore, this cut is applied to all sources with $z_\mathrm{p}<0.9$.

\subsection{The DESI-Calibrated HSC Y3 $n(z)$}
\label{sec:data:desi}

This work presents updated cosmic shear constraints using the calibrated HSC Y3 $n(z)$ distributions obtained in the companion paper \cite{ChoppinDeJanvry2025a}. In that work, we use DESI Data Release 1 (DR1) and DR2 to recover the redshift distribution of the four aforementioned tomographic bins by leveraging the spatial clustering of galaxies via the clustering redshifts method \cite{Newman2008ClusterZ, HSCClusteringRau2023, EuclidClusterRedshifts2025, Schmidt2013ClusteringRedshifts, ménard2014clusteringbasedredshiftestimationmethod}. 
The Dark Energy Spectroscopic Instrument \cite{DESI.DR1.I.Presentation} is an eight year multiplexed spectroscopic program \cite{DESI2016b.Instr, DESI.DR2.II.BAO} started in 2021, conducting a survey of $17,000$ deg$^2$ of the sky, covering spectroscopic redshift ranges up to $z\sim4.2$. As such, it provides an unprecedented reference spectroscopic dataset to calibrate photometric redshifts. 
We measure the angular cross-correlation between the HSC Y3 photometric sample and the DESI spectroscopic reference samples as a function of the spectroscopic redshift. Since galaxies at similar redshifts are physically clustered together, the cross-correlation signal peaks when the photometric and spectroscopic samples overlap in true redshift space. 

\begin{figure}[ht]
    \centering
    \includegraphics[width=0.95\textwidth]{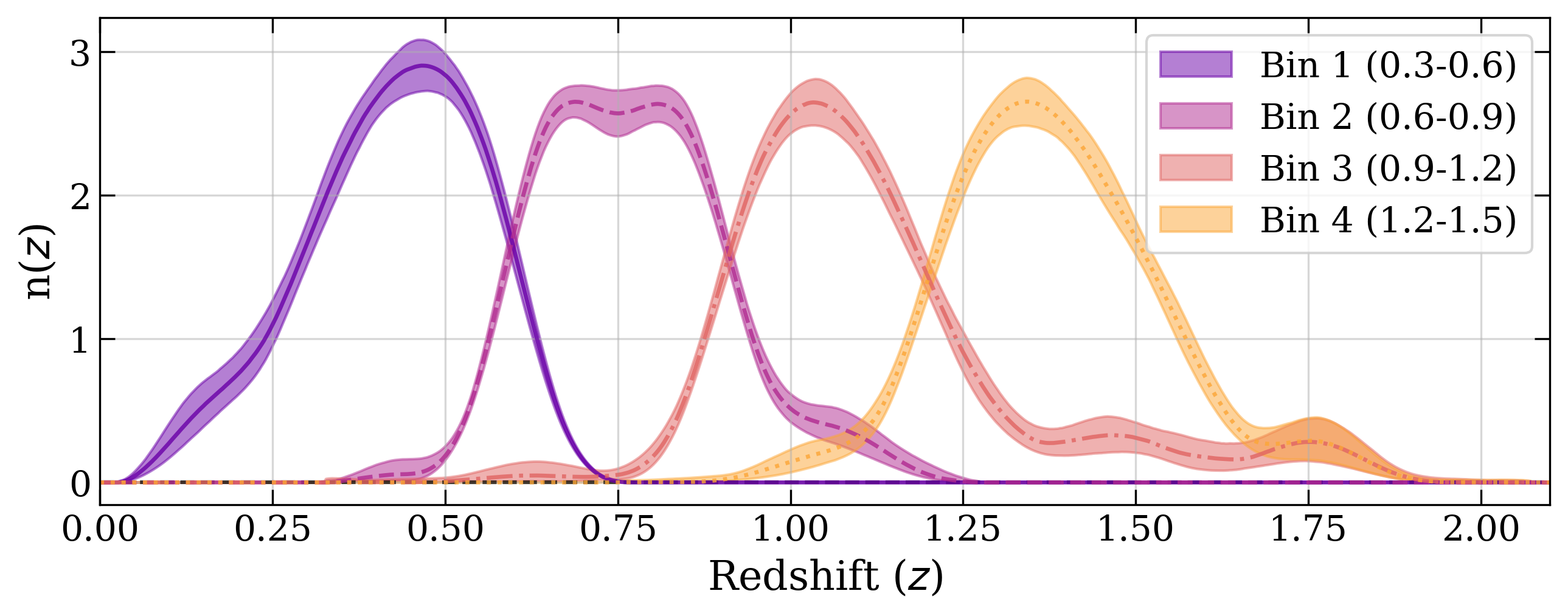}
    \caption{$n(z)$ distributions for each of the 4 tomographic Bins, spanning $z_\mathrm{p}\sim0.3$ to $z_\mathrm{p}\sim1.5$ where $z_\mathrm{p}$ is the best estimate of the \texttt{DNNz} photo-$z$ algorithm \cite{PhotoZHSCNishizawa} used for tomographic Bin selection, as described in section \ref{sec:data:hsc}. The $n(z)$ for each Bin is displayed with the median posterior sample and the $1\sigma$ widths adapted from \cite{ChoppinDeJanvry2025a}. Here, the $n(z)$ shown includes all corrections to galaxy bias and magnification effects. The $n(z)$ displayed here is inferred with the $[0.3,\,3]\hMpc$ scale cut.}
    \label{fig:all_tomo}
\end{figure}

We analyze how the cross-correlation amplitude varies with spectroscopic redshift, characterize and model systematic biases (including galaxy bias from both the spectroscopic and photometric samples and magnification effects) and finally, we reconstruct a fully DESI-calibrated underlying redshift distribution $n(z)$ of the four tomographic bins of the HSC Y3 photometric sample, displayed in Figure \ref{fig:all_tomo}. The distributions are modeled from the data points using a basis of splines. In the clustering redshifts analysis, we include two scale cuts used for two-point correlation functions in the $n(z)$ calibration measurement ($0.3-3$h$^{-1}$Mpc and $1-5$h$^{-1}$Mpc). The smaller scale cut is the fiducial measurement of this work: however, we make use of both scale cuts for consistency purposes, as presented in section \ref{sec:results}. For more details on the methodology followed, we refer the reader to Choppin de Janvry {\it et al.} (2025a) \cite{ChoppinDeJanvry2025a}.

%% file: sections/method.tex
\section{Method}
\label{sec:method}

This section presents the methodology followed in this analysis to recover cosmological constraints. The constraints are inferred with the following methods: (a) in Section~\ref{sec:method:mcmc}, we re-run the likelihood analysis following the HSC real space cosmic shear analysis \cite{Li2023_HSCY3_CosmicShear}, using the updated redshift distributions and redshift parameter priors; (b) in Section~\ref{sec:method:importance}, we apply importance weighting to the redshift distribution parameters of the fiducial HSC Y3 cosmic shear chains \cite{Li2023_HSCY3_CosmicShear,Dalal2023CosmoShearHSC} based on the updated redshift parameter priors. The latter serves as a mean of inexpensive inference based on the original chains. Finally, we investigate some additional systematics to assess the robustness of our distributions in Section~\ref{sec:method:systematic}.

\subsection{Likelihood Analysis}
\label{sec:method:mcmc}

The fiducial towards cosmological parameter inference is to re-run the HSC Y3 fiducial likelihood analysis with our DESI-calibrated redshift distribution $n(z)$ for all four tomographic bins. 

The real-space cosmic shear likelihood analysis follows the setup presented in \cite{Li2023_HSCY3_CosmicShear} as close as possible, which enables us to highlight the effects caused solely by the photometric redshift calibration. This analysis setup follows the modeling choices in \cite{Li2023_HSCY3_CosmicShear} in terms of the calculation of the baryonic physics, galaxy intrinsic alignment, shear multiplicative bias, and PSF additive bias. 
For all parameters besides the photo-$z$ parameters, we follow the choices performed in \cite{Li2023_HSCY3_CosmicShear, Dalal2023CosmoShearHSC} and implement a total of 23 free parameters. 
In particular, $n(z)$ marginalization is done through a single parameter per bin, a mean shift $\Delta z_{i}$ with $i\in\llbracket 1\,, 4\rrbracket$. For HSC Y3, \cite{Zhang2022PhotometricRedshiftShifts} has shown such modeling is sufficient to capture the variance in $n(z)$ estimates, and therefore we keep the same scheme for a faithful comparison. However, future lensing surveys should consider more complex marginalization schemes, for example \texttt{HYPERRANK} and \texttt{MULTIRANK} \cite{Cordero2022_marginalization_nz} used for DES Y3. 

The cosmological inference pipeline is implemented in the public software \texttt{CosmoSIS}\footnote{\href{https://github.com/cosmosis-developers/cosmosis-standard-library}{\textcolor{blue}{github.com/cosmosis-developers/cosmosis-standard-library}}} and the configuration files used in this analysis are made available through the common repository for this work and the companion $n(z)$ calibration analysis \citep{ChoppinDeJanvry2025a}: \href{https://github.com/JeanCHDJdev/desi-y3-hsc}{\textcolor{blue}{github.com/jeanchdjdev/desi-y3-hsc}}. The priors for all of the 23 parameters are described in table \ref{tab:post}. The fixed parameters are given in table \ref{tab:fixed} in appendix \ref{sec:appendix}.
For the matter power spectrum, we use the CAMB\footnote{\href{https://camb.info/}{\textcolor{blue} {camb.info}}} \cite{LewisCAMB2011} linear power spectrum emulator, in a divergent approach from using the BACCO suite of emulators  \cite{Arico2022BACCOEmulator} \footnote{\href{https://baccoemu.readthedocs.io/en/latest/}{\textcolor{blue}{baccoemu.readthedocs.io}}} used in \cite{Li2023_HSCY3_CosmicShear}. It is shown that there is no significant different in cosmological results between power spectrum computation with CAMB and emulation with the BACCO emulator for HSC Y3 \cite{Li2023_HSCY3_CosmicShear} in $S_8$, as demonstrated in Fig 8 (contained to $\sim0.15\sigma$). However, biases on $\Omega_\mathrm{m}$ are observed, especially on the projected 1D mode, ranging from $-0.4\sigma$ to $-0.8\sigma$. Further discussion on power spectrum computation is discussed in \cite{ferri2026contributionsmallscalestwopoint}, in the context of HSC Y3. Due to these potential biases, we follow the approach led by the original cosmic shear analysis and focus the inferred constraints on $S_8$. 

The two-point correlation functions of galaxy shear are denoted $\xi_\pm$ \cite{Li2023_HSCY3_CosmicShear} and are measured on the shear catalog introduced in section \ref{sec:data:hsc}.
We also follow the same formalism when it comes to likelihood calculation, compared to \cite{Li2023_HSCY3_CosmicShear}.
The scale cut applied in this analysis is the same as  \cite{Li2023_HSCY3_CosmicShear}: $7.1 < \theta < 56.6$ [arcmin] for $\xi_+(\theta)$, and $31.2 < \theta < 248$ [arcmin] for $\xi_+(\theta)$. We adopted the same covariance matrix calculated by 1404 mock HSC Y3 catalogs, based on mocks developed by \cite{TakahashiMocksHSC}. We note that this is an approximation, since the covariance matrix uses information from the original $n(z)$, and we change the $n(z)$ used in this analysis. The covariance modification due to this change is neglected. We also apply the same Hartlap factor \cite{Hartlap2008} to the likelihood calculation to account for covariance matrix underestimation due to limited mock catalogs. In the HSC Y3 fiducial pipeline, the $S_8$ parameter is not sampled and is instead a computed property from the other sampled parameters. 

The sampler we used for the likelihood analysis is \texttt{pocoMC}\footnote{\href{https://pocomc.readthedocs.io/en/latest/}{\textcolor{blue}{pocomc.readthedocs.io}}} \cite{karamanis2022pocoMC1, karamanis2022pocoMC2}. \texttt{pocoMC} utilizes the Preconditioned Monte Carlo (PMC) algorithm which offers significant speed-ups over more traditional approaches such as Markov Chain Monte Carlo (MCMC). We verify \texttt{pocoMC} accurately recovers the fiducial HSC chains in \texttt{CosmoSIS}. 2D posteriors for parameters are obtained with the \texttt{getDist}\footnote{\href{https://getdist.readthedocs.io/en/latest/}{\textcolor{blue}{getdist.readthedocs.io}}} \cite{getDistLewis2025} package. Comparatively to \cite{Li2023_HSCY3_CosmicShear}, this work does not use the \texttt{ChainConsumer}\footnote{\href{https://samreay.github.io/ChainConsumer/}{\textcolor{blue}{samreay.github.io/ChainConsumer/}}} \cite{ChainConsumer_Hinton2016} package to treat the chains, due to potential biases incorporated by the boundary effects and smoothing of MC samples at the border with Kernel Density Estimation (KDE) mentioned in Appendix A of \cite{Li2023_HSCY3_CosmicShear}. However, constraints are also computed with \texttt{ChainConsumer} and consistency with \texttt{getDist} is confirmed for parameters $S_8$ and $\Omega_\mathrm{m}$. Posterior distributions for a few parameters using both \texttt{getDist} and \texttt{ChainConsumer} are shown in appendix \ref{sec:appendix:chain_processing}. 

In this work, the posterior of cosmological parameter constraints is presented in the form of:
\begin{equation}\label{eq:presentation}
    \text{1D mode} ^ {+34\%} _ {-34\%}\;\text{(MAP)}
\end{equation}
where the 1D mode is the is 1D mode of the marginalized 1D posterior distribution, the MAP is the Maximum A Posteriori point of the chain (i.e. the best fit model, including the priors), and the error interval is the asymmetric $34\%$ confidence interval around the mode of the distribution, computed with the cumulative density function.
A large deviation between the 1D mode and the MAP is often an indication of prior volume projection effects caused by fitting too many parameters that are not directly observable. The values themselves are presented as obtained by \texttt{getDist} using \texttt{MCSamples.get1DDensity} and the weights are obtained through the sampling from \texttt{pocoMC} or included in the fiducial HSC Y3 chains.

\subsection{Importance Weighting}
\label{sec:method:importance}

A secondary approach is to apply importance weights to the photometric redshift shifts $\Delta z$ on the original chains \cite{Li2023_HSCY3_CosmicShear}. This approach serves as an additional sanity test and is not representative of our fiducial results, which are more accurate by MCMC analysis. Furthermore, the resampling affects edges of the posterior volume: this makes for extreme weights and difficult smoothing of the posterior distribution after reweighting, especially given the lower amount of representative samples. The importance weights are only applied to the last two tomographic bins, leveraging the broad uniform ($\mathcal{U}(-1,1)$) that were applied to the shift parameters in the original analysis. No resampling is applied to the first two bins because of their narrow gaussian prior already set by original HSC analysis, which would lead to extreme weights due to the ratio of the gaussian distributions. Furthermore, Bin 1's new calibration presents a slight shift compared to the original $n(z)$ presented in \cite{HSCClusteringRau2023}, but does not offer significant cosmological information due to being at low redshifts ($0.3\lesssim z_\mathrm{p}\lesssim0.6$), so it is possible to neglect reweighting in Bin 1. Bin 2's calibration is fully consistent with the calibration presented by shear ratios (SR) \cite{Rana2025HSCY3CosmicShearRatios} and the original calibration \cite{HSCClusteringRau2023}.

The obtained result is also consistent with the sampling result, as shown in section \ref{sec:results}, however, the posterior volume is larger, attributed to larger priors on the first two Bins, since the reweighting only affects bins with the uniform prior in the original analysis. The weights $w^{(k)}_i$ derived for Bin $k\in\{3,\,4\}$ are:
\begin{equation}
    w^{(k)}_i=\mathrm{exp}\left(-\frac{1}{2}{\left(\frac{\mu^{(k)}_i-\mu^{(k)}}{\sigma^{(k)}}\right)^2}\right)
\end{equation}
where $\mu^{(k)}_i$ is the mean redshift shift of Bin $k$ of posterior draw $i$ from the original chain, $\mu^{(k)}$, $\sigma^{(k)}$ are respectively the mean and standard deviations of the new prior. Since no weighting is originally applied from the uniform prior, we can multiply the existing weights in MCMC chains with $w^{(k)}_i$. 

\subsection{Systematics}\label{sec:method:systematic}

In this section, we investigate the robustness of the $n(z)$  distributions used. First, magnification coefficients are assumed to have errors that systematically perturb the magnification-corrected distributions. Second, we assume a polynomial model instead of a power law for the bias evolution correction term of the photometric sample. We show both of these effects are negligible effects on our error budget. 

\paragraph{Magnification error propagation} In the fiducial analysis \cite{ChoppinDeJanvry2025a}, magnification bias coefficients ($\alpha_p$,$\alpha_s$) had no error associated were not marginalized into the distributions. Here we adopt a "systematic perturbation" shift, drawn from $\mathcal{N}(0,\sigma_\alpha=0.1)$ for each $\alpha_p$. We note that this shift is larger than the reported errors for $s_\mu$ given in \cite{ChoppinDeJanvry2025a} (for the coefficients that report errors). This perturbs the value of $\alpha_p$ and $\alpha_s$ independently of redshift. We perform 100 realizations and measure the variance of the expectation from these measurements, after re-fitting the splines to the modified measurements. The variance is appended to the statistical budget.

\paragraph{Modeling of the photometric sample bias evolution} Another modeling choice was using a power-law model for the proxy to the bias evolution of the photometric redshift sample, $\sqrt{\bar\omega_{pp}(z)}$. This is further detailed in Section 3.3 of \cite{ChoppinDeJanvry2025a}. In \cite{EuclidClusterRedshifts2025}, a $2^\mathrm{nd}$ order polynomial $az^2+bz+c$ was used, with three free parameters. We follow this choice and fit $\sqrt{\bar\omega_{pp}(z)}$ to the polynomial. In each bin, this leads to $\delta z$ shift in the mean. We include this in the total budget by summing the variance. \\

\begin{figure}
    \centering
    \includegraphics[width=0.98\linewidth]{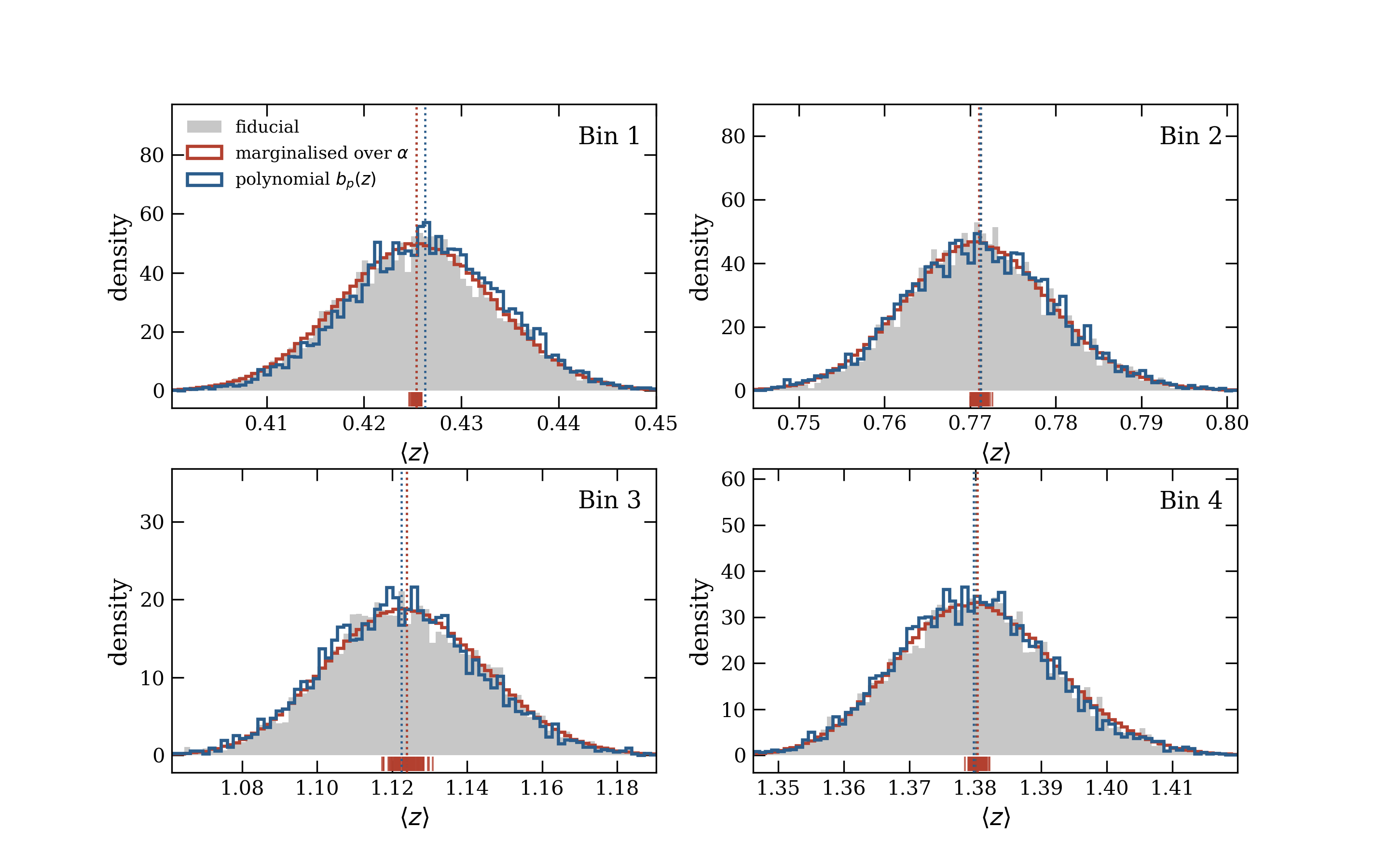}
    \caption{Posterior distributions for the systematic effects In red, we display the contribution of perturbations to the magnification. The expectation values for each of the 100 $\alpha$ draws are shown in red, below the histogram. The scatter of this expectation is named $\sigma_\mu$ (following notations from \cite{ChoppinDeJanvry2025a}) is added as systematic variance to our priors. The bias due to the polynomial fit versus the power law are shown in blue. The mean difference $\delta z$ is squared and also included in our priors.}
    \label{fig:syst}
\end{figure}

Therefore, the total systematic budget we add to the run is
$\sigma_{tot}=(\sigma_\mathrm{stat}^2+\sigma_\mu^2+(\delta z)^2)^{1/2}$ per each tomographic bin. Figure~\ref{fig:syst} displays the posterior distributions with the aforementioned modifications. We conclude these two systematics are minor with respect to the current statistical budget. In order to include the systematic budget into the run, we first run the chains with the fiducial priors (composed only of $\sigma_\mathrm{stat}$ derived priors). Then, we re-weight the chains to broaden the priors, using the method described in \ref{sec:method:importance}, such that weights are relative (maximum weight is 1). Here, taking the ratio of Gaussian priors is not an issue because they are centered on similar values with similar widths. We verify the Effective Sample Size (ESS) remains large enough for accurate posteriors after re-weighting.
The increase in prior size is negligible ($\sim0.5-2\%$ approximately). The largest increase is in Bin 3: this is expected, since it has the largest spread over redshift and therefore can have larger magnification effects.

\paragraph{Scale cut differences} The reported expectation values $\bar{z}_i$ for each scale cuts, presented in \cite{ChoppinDeJanvry2025a}, differ slightly. We show in this paragraph that such differences can be attributed to statistical differences and do not require an introduction of an additional systematic error. Assuming the scales $3-5\hMpc$ do not significantly contribute to information, the scale cuts essentially compare $1-3\hMpc$ to $0.3-3\hMpc$. This is helped by the weighting function $W(r)$ applied during integration of the two-point function measurements, as described in \cite{ChoppinDeJanvry2025a}, where higher scales are down-weighted. The standard deviation of the parameter shift is  $\sigma({\bar z_4})=\sqrt{\sigma_{0.3-3}^2-\sigma_{1-5}^2}\simeq0.106$, with $\sigma_{0.3-3}=0.012$ and $\sigma_{1-5}=0.016$ the error bar of the respective scale cuts. The parameter shift for $\Delta z_4$ between the two scale cuts is therefore fully compatible within $1\sigma$. Furthermore, including extra information from 3-5$\hMpc$ will only increase the standard deviation of the parameter shift, so $\sigma({\bar z_4})$ is a lower bound to the true parameter shift standard deviation. We conclude there is no sign of additional systematics with respect to the scale cuts.

In recent years, weak lensing surveys \cite{DES2025_ClusteringSOM, HSCClusteringRau2023, wright2025kidslegacyClusterZ} have opted for varied methods to estimate their $n(z)$ distributions. In particular, popular methods focus on machine learning or importance reweighting of spectroscopic datasets to calibrate the color-space of the observed photometric sample \cite{lange2025unifiedphotometricredshiftcalibration, SOMPZDES2024, wright2025kidslegacyClusterZ, DES2025_ClusteringSOM} and combine this information with clustering redshifts. In this work, the $n(z)$ distribution of each tomographic bin is solely derived from clustering redshifts, as no reliable independent calibration is currently available for all HSC tomographic bins. We do not combine our information with the developed \textbf{PhotZ} described in \cite{HSCClusteringRau2023}. The primary reason is that \textbf{PhotZ} alone was deemed unreliable at high ($z\gtrsim1.2$) redshifts by the cosmic shear analyses \cite{Dalal2023CosmoShearHSC, Li2023_HSCY3_CosmicShear}, which in turn preferred relying on the self calibration approach.  For the lowest redshift bins, we report good agreement, 
with a mild discrepancy at $z\sim0.2$ akin to the offset reported by \cite{HSCClusteringRau2023} between their clustering measurements and \textbf{PhotZ}. The second redshift bin shows one measurement that has a mild offset with respect to \textbf{PhotZ} at $z\sim1.1$, where spectroscopic calibration was not readily available for \textbf{PhotZ}. 
Finally, newly developed validation methods such as direct calibration with DESI redshifts \cite{lange2025unifiedphotometricredshiftcalibration, abby2026_inprep} can be applied to HSC Y3 and other recent data releases of weak lensing catalogs, which will provide an additional 
validation of our results. 

%% file: sections/results.tex
\section{Results}
\label{sec:results}

In this section, we present the fiducial constraints we obtain from the weak lensing cosmological inference in real space and compare them to other experiments and surveys. Measurements from this analysis are compared to other measurements from Stage III weak lensing surveys and cosmic microwave background (CMB) measurements, assessing the relevance of the $S_8$ tension in light of these new measurements. In this work, following \cite{Li2023_HSCY3_CosmicShear, Dalal2023CosmoShearHSC}, the constraint on the $S_8$ parameter is emphasized, due to the strong degeneracies between parameters $\Omega_\mathrm{m}$ and $\sigma_8$.

Furthermore, we perform sanity checks of this analysis by comparing the results to importance sampling of the fiducial chains, as well as using different configuration for the tomographic $n(z)$ distributions. We note that following the configuration set in \texttt{CosmoSIS}, the convention used for $\Delta z$ parameter shifts is opposed to the one presented in the clustering redshift analyses \cite{HSCClusteringRau2023, ChoppinDeJanvry2025a} and in the cosmological inference works such as \cite{Li2023_HSCY3_CosmicShear, Dalal2023CosmoShearHSC, Zhang2025aTomographic2x2pt, Zhang2025bTomographic3x2pt}. Here, a positive shift implies that the real $n(z)$ distribution is at higher redshifts than the original, calibrated one proposed in the redshift calibration analyses: the original convention implied the same for negative shifts.

\begin{figure}[ht]
    \centering
    \includegraphics[width=0.95\textwidth]{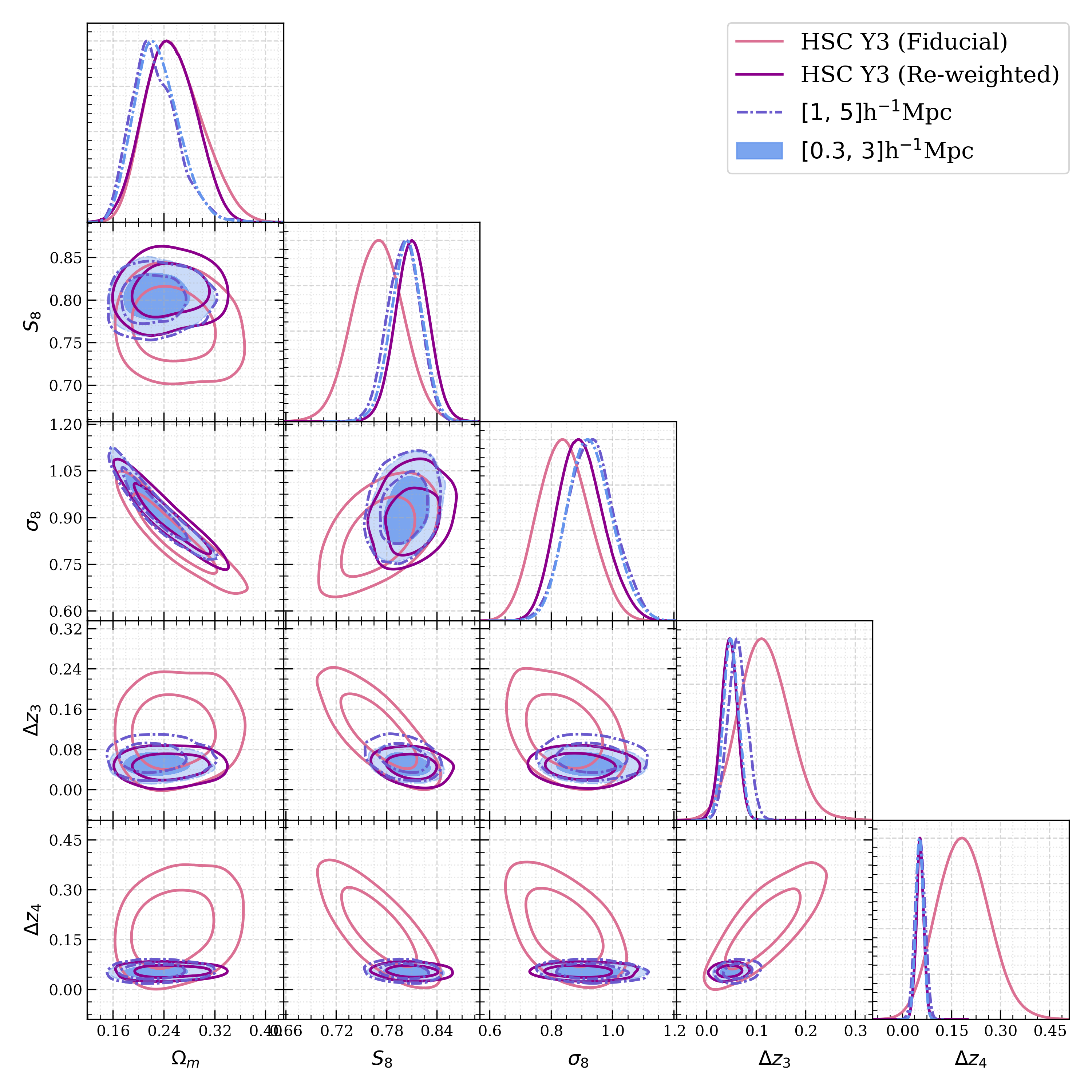}
    \caption{Comparing the sampled results (section \ref{sec:method:mcmc}) from the two $n(z)$ set of distributions at different scale cuts with the importance re-weighting (section \ref{sec:method:importance}) approach in real space and to  the HSC Y3 fiducial result. The shift posteriors are artificially shifted by the expectation of the inferred shift in \cite{ChoppinDeJanvry2025a}, as they are by default centered around 0 for the sampling approach, since new $n(z)$ distributions are used.}
    \label{fig:corner_resampling_sc}
\end{figure}

Figure \ref{fig:corner_resampling_sc} compares the joint 2D posteriors for the fiducial result, the $1-5\hMpc$ scale cut $n(z)$ result to the importance sampling in real space and the fiducial HSC Y3 result. The underlying $\Delta z_{\{3,\,4\}}$ shifts lie at the edge of the reported posterior from HSC Y3. While the shifts are significant in value, they are not statistically incoherent with the self-calibration method deployed in HSC Y3. This implies the resampling approach will propose extreme weights due to the new distributions lying at the edge of the posterior space, hence providing fairly unstable posterior distributions with less representative samples.

\begin{figure}[ht]
    \centering
    \includegraphics[width=0.95\textwidth]{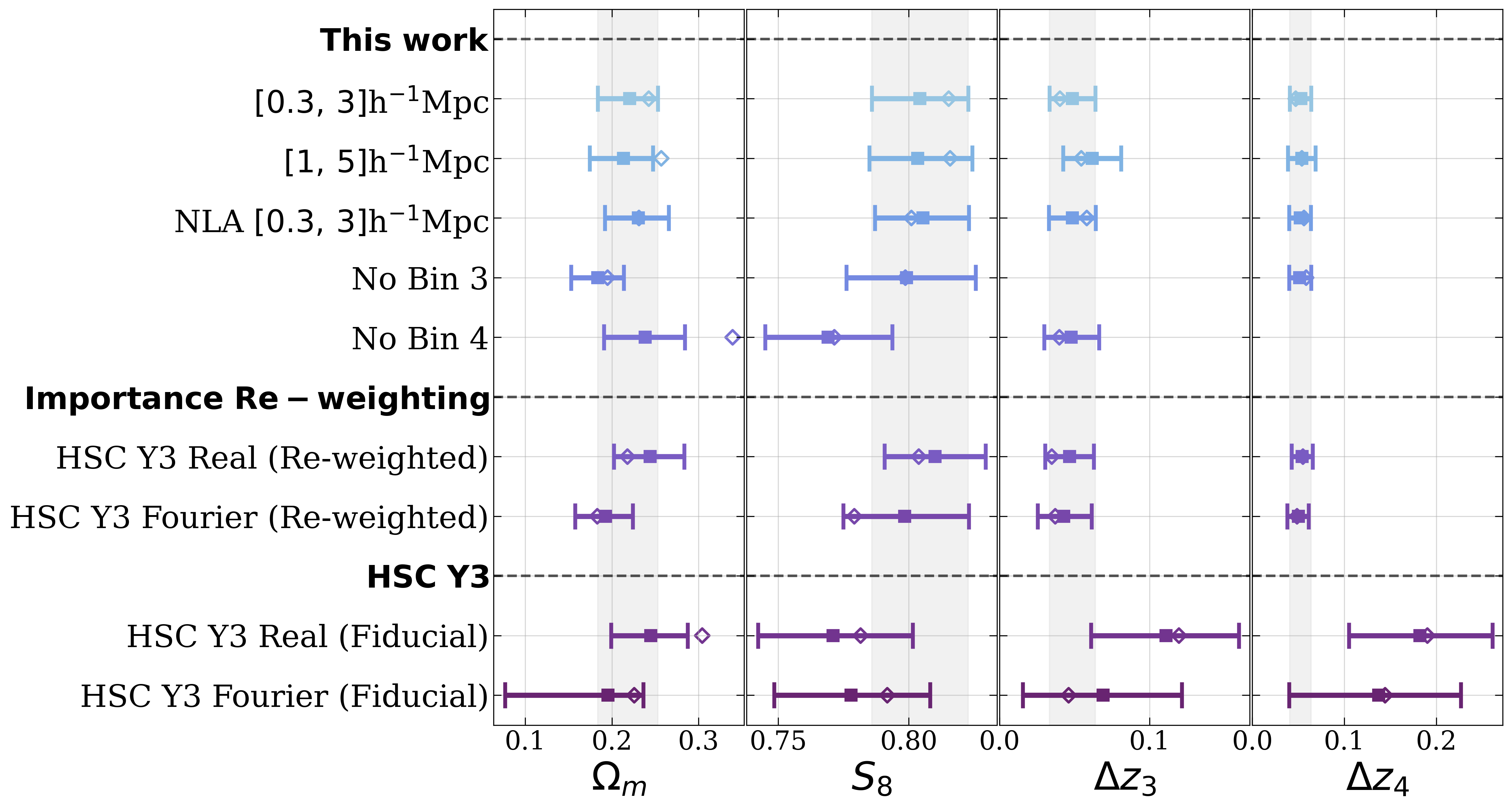}
    \caption{$\Omega_\mathrm{m}$, $S_8$, $\Delta z_3$, $\Delta z_4$ parameters under different analysis choices, and compared to the original HSC Y3 results. This includes the sampling approach, with the $n(z)$ distributions at the two different scale cuts. The measurements obtained when excluding either Bin 3 or Bin 4 or using non-linear intrinsic alignment (NLA) modeling are shown. These are then put in context of the importance re-weighting and fiducial results of the Real and Fourier HSC Y3 analyses \cite{Li2023_HSCY3_CosmicShear, Dalal2023CosmoShearHSC}. We note that the reported differences between re-weighting and MCMC are interesting to compare in real-space specifically, since the fiducial results of this work come from the likelihood inferred in real-space. The square marker displays the 1D posterior mode obtained with \texttt{getDist}. The diamond marker is the Maximum A Posteriori (MAP) value. The background gray area represents the projected fiducial measurement obtained with the $0.3-3\hMpc$ scale cut for the $n(z)$. The $n(z)$ distributions are not the same for HSC Y3 samples (derived from \cite{HSCClusteringRau2023}) and this work (from \cite{ChoppinDeJanvry2025a}) : the posteriors of this work have been artificially shifted to match the mean of the shifts found in \cite{ChoppinDeJanvry2025a}.}
    \label{fig:param_sigma_shifts}
\end{figure}

Figure \ref{fig:param_sigma_shifts} presents stress tests to the analysis and compare the reported posterior value. 
The importance re-weighting results obtained with the method described in section \ref{sec:method:importance} are shown next. 
The shift in the $\Delta z$ bins in the context of the importance re-weighting, the $1-5\hMpc$ distributions or in the context of the HSC Y3 fiducial $n(z)$ derived in \cite{HSCClusteringRau2023} cannot be strictly compared to the $n(z)$ shifts found in this work, since the $n(z)$ distributions themselves differ. 
Lastly, the importance re-weighting of the HSC Y3 chains means constraining the originally broad uniform $\mathcal{U}(-1, 1)$ prior over $\Delta z_3$, $\Delta z_4$ to a stricter gaussian prior. Reweighting leads to a fairly low effective sample size with respect to the original chain, leading to noisy 2D posteriors, as specified in Section \ref{sec:method:importance}. 
Thus, the MCMC results are considered as the fiducial results of this work. 

We also investigate Intrinsic Alignment (IA) differences. The DES Y6 cosmic shear \cite{DESy6_S8_CosmicShear} results show significant shifts between the NLA (Non-Linear Alignment \cite{Hirata_2007_IA_NLA}) and TATT (Tidal Alignment and Tidal Torque \cite{Blazek_2019_IA_TATT}) models of intrinsic alignment. The implementation of TATT IA in \texttt{CosmoSIS} uses the \texttt{FAST-PT} \cite{McEwen_2016_FASTPT, Fang_2017_FASTPT} algorithm to rapidly compute to one-loop order coupled contributions to the power spectra contributions for II (intrinsic shape-intrinsic shape) and GI (galaxy lensing-intrinsic shape) correlations. In this analysis, we use 5 free parameters ($A_1$, $\eta_1$, $A_2$, $\eta_2$, $b_{\mathrm{TA}}$) for TATT. NLA follows the same setup, but restrict $A_2$, $b_{\mathrm{TA}}$, $\eta_2$ to 0. We refer to the HSC Y3 cosmic shear analyses \cite{Li2023_HSCY3_CosmicShear, Dalal2023CosmoShearHSC} for a more thorough description of the above parameters. Both are implemented directly in \texttt{CosmoSIS}. In this work, we report the same trend as the HSC Y3 cosmic shear analyses, showing no significant differences between adopting TATT or NLA for IA. We conserve TATT as the fiducial model choice, in order to remain consistent with HSC Y3 choices. It is nonetheless interesting to compare the DES Y6 results in NLA, showing good agreement with HSC Y3. On the contrary, the TATT model for DES Y6 shows slightly more bias with respect to HSC Y3, due to the differences reported by the cosmic shear analysis of DES Y6 \cite{DESy6_S8_CosmicShear}.

\begin{figure}[ht]
    \centering
    \includegraphics[width=0.7\textwidth]{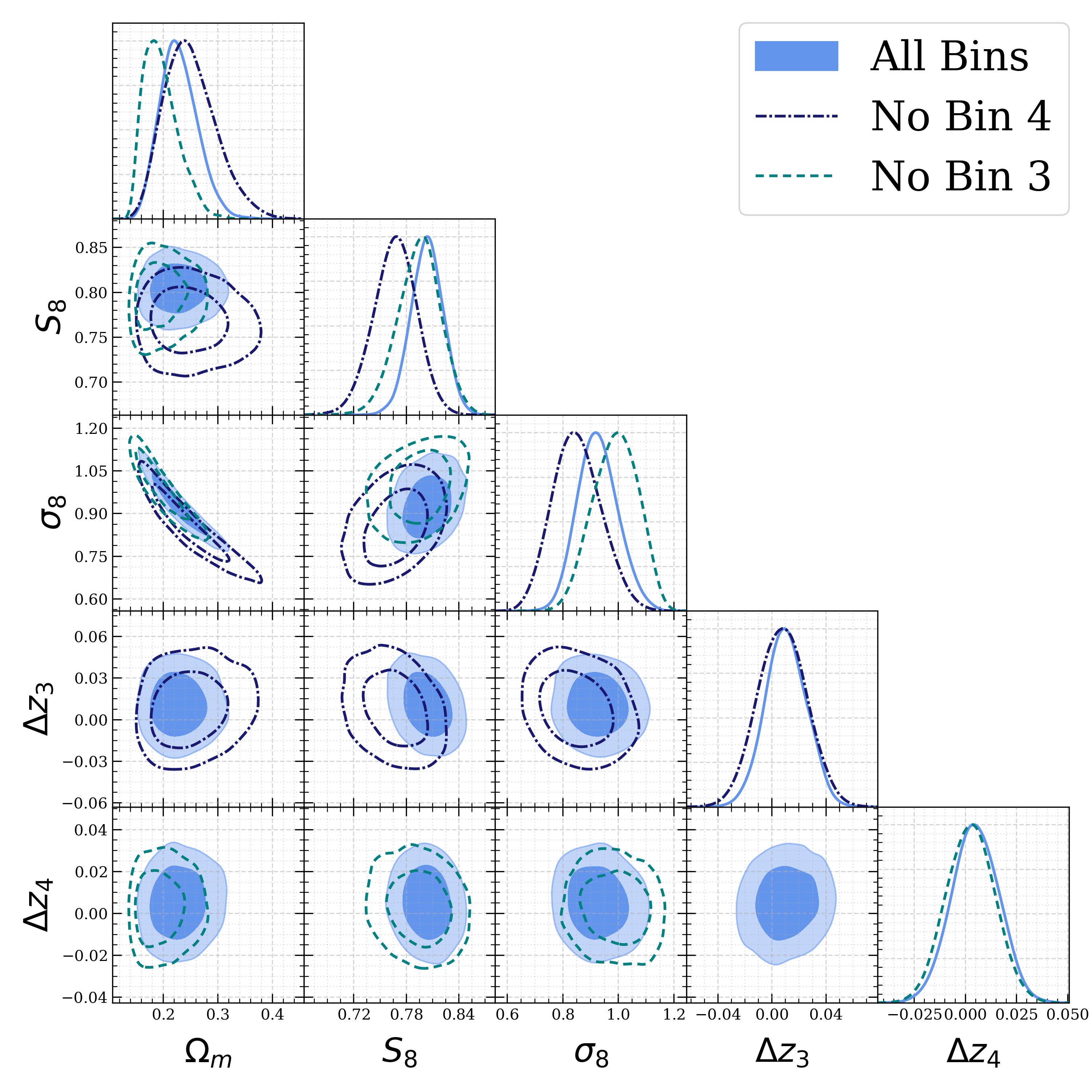}
    \caption{To assess the robustness of the measurement, we exclude Bin 3 and Bin 4 respectively from the chain inference. The measurements are shown to have decent overlap between the two and with the "All Bins" measurement.}
    \label{fig:corner_excluded_bins}
\end{figure}

We consider parameter estimation when excluding bins from MCMC inference, presented as the "No Bin \textit{i}" results.
The lower error bound of the "No Bin 3" chain is underestimated due to the MCMC chain piling up near the boundaries of the prior ($\pi(\Omega_\mathrm{m})\sim\mathcal{U}(0.1, 0.7)$). 
Furthermore, the obtained value when excluding Bin 3 presents a fairly low value of $\Omega_\mathrm{m}$, akin to behavior presented in Figure 17 of the two-point correlation analysis \cite{Li2023_HSCY3_CosmicShear}. Nonetheless, the bin exclusion results overlap well with the "All Bins" results. 
In Figure \ref{fig:corner_excluded_bins}, we present the corner plots associated to the aforementioned measurements to compare the posterior distribution overlaps.

{\renewcommand{\arraystretch}{1.2}
\begin{table}[ht]
\centering
\caption{\label{tab:post}
Fiducial parameters and priors for this cosmological analysis. The values are presented according to equation \ref{eq:presentation}. The redshift shifts are relative to our derived distributions, and are centered around zero. Note that baryonic feedback 
posterior distribution is very non-Gaussian and 
posterior density has non zero support at zero feedback value of 
$A_b=3.13$. 
}
\begin{tabular}{l | l | l | l }
\hline
\textbf{Parameter} & \textbf{Prior} & $\mathbf{[0.3,\,3]}$h$\mathbf{^{-1}}$\textbf{Mpc} & $\mathbf{[1,\,5]}$h$\mathbf{^{-1}}$\textbf{Mpc} \\ 
\hline
\hline
\textbf{Cosmological Parameters} &  & & \\
$A_s\,(\times 10^9)$ & $\mathcal{U}(0.5, 10)$ & $3.721_{-1.820}^{+1.284} ({2.765})$ & $3.730_{-2.092}^{+1.371} ({2.922})$ \\ 
$\Omega_\mathrm{m}$ & $\mathcal{U}(0.1, 0.7)$ & $0.221_{-0.037}^{+0.033} ({0.243})$ & $0.214_{-0.039}^{+0.034} ({0.257})$ \\ 
$\sigma_8$ & -- & $0.922_{-0.071}^{+0.073} ({0.907})$ & $0.936_{-0.071}^{+0.083} ({0.882})$ \\ 
$S_8$ & -- & $0.805_{-0.018}^{+0.018} ({0.815})$ & $0.804_{-0.019}^{+0.020} ({0.816})$ \\ 
$n_s$ & $\mathcal{U}(0.87, 1.07)$ & $0.900_{-0.041}^{+0.053} ({0.885})$ & $0.881_{-0.023}^{+0.049} ({0.898})$ \\ 
$h_0$ & $\mathcal{U}(0.62, 0.80)$ & $0.662_{-0.056}^{+0.052} ({0.746})$ & $0.701_{-0.052}^{+0.059} ({0.724})$ \\ 
$\omega_b \equiv \Omega_b h^2$  & $\mathcal{U}(0.02, 0.025)$ & $0.024_{-0.002}^{+0.003} ({0.020})$ & $0.020_{-0.000}^{+0.002} ({0.024})$ \\ 
\hline
\textbf{Baryonic Feedback} & & & \\
$A_{\rm b}$ & $\mathcal{U}(2, 3.13)$ & $2.075_{-0.128}^{+0.243} ({2.083})$ & $2.041_{-0.089}^{+0.226} ({2.137})$ \\ 
\hline
\textbf{Intrinsic Alignment} & & & \\
$A_1$ & $\mathcal{U}(-6, 6)$ & $0.441_{-0.334}^{+0.356} ({0.648})$ & $0.432_{-0.373}^{+0.369} ({0.503})$ \\ 
$\eta_1$ & $\mathcal{U}(-6, 6)$ & $-2.478_{-2.553}^{+2.884} ({-3.549})$ & $-1.943_{-2.717}^{+2.853} ({-4.323})$ \\ 
$A_2$ & $\mathcal{U}(-6, 6)$ & $-0.058_{-0.840}^{+0.748} ({-0.603})$ & $-0.057_{-0.841}^{+0.778} ({-0.469})$ \\ 
$\eta_2$ & $\mathcal{U}(-6, 6)$ & $-1.524_{-3.149}^{+3.143} ({-3.915})$ & $-1.552_{-3.511}^{+3.048} ({-4.782})$ \\ 
$b_{\rm TA}$ & $\mathcal{U}(0, 2)$ & $0.062_{-0.135}^{+0.421} ({0.061})$ & $0.067_{-0.159}^{+0.441} ({0.163})$ \\ 
\hline
\textbf{Redshift shifts} & & & \\
$\Delta z_1$ & $\mathcal{N}(0, 0.008)$ & $-0.002_{-0.008}^{+0.008} ({-0.002})$ & $-0.003_{-0.013}^{+0.013} ({-0.004})$ \\ 
$\Delta z_1$ & $\mathcal{N}(0, 0.009)$ & $-0.002_{-0.008}^{+0.009} ({-0.009})$ & $-0.010_{-0.011}^{+0.012} ({-0.027})$ \\ 
$\Delta z_3$ & $\mathcal{N}(0, 0.020)$ & $0.007_{-0.016}^{+0.015} ({0.001})$ & $0.020_{-0.019}^{+0.019} ({0.015})$ \\ 
$\Delta z_4$ & $\mathcal{N}(0, 0.012)$ & $0.002_{-0.012}^{+0.011} ({-0.001})$ & $0.005_{-0.015}^{+0.015} ({0.005})$ \\ 
\hline
\textbf{Shear calibration} & & & \\
$\Delta m_1$ & $\mathcal{N}(0, 0.01)$ & $0.000_{-0.010}^{+0.011} ({-0.009})$ & $0.001_{-0.010}^{+0.011} ({-0.003})$ \\ 
$\Delta m_2$ & $\mathcal{N}(0, 0.01)$ & $-0.003_{-0.009}^{+0.010} ({0.001})$ & $-0.005_{-0.010}^{+0.010} ({-0.003})$ \\ 
$\Delta m_3$ & $\mathcal{N}(0, 0.01)$ & $0.000_{-0.009}^{+0.009} ({0.001})$ & $0.001_{-0.010}^{+0.009} ({-0.001})$ \\ 
$\Delta m_4$ & $\mathcal{N}(0, 0.01)$ & $0.005_{-0.010}^{+0.009} ({0.010})$ & $0.007_{-0.009}^{+0.011} ({0.005})$ \\ 
\hline
\textbf{PSF systematics} & & & \\
$\tilde{\alpha}_2$ & $\mathcal{N}(0, 1)$ & $0.075_{-0.937}^{+0.997} ({0.159})$ & $-0.094_{-1.093}^{+0.922} ({-0.109})$ \\ 
$\tilde{\beta}_2$ & $\mathcal{N}(0, 1)$ & $0.033_{-0.965}^{+0.939} ({0.160})$ & $0.102_{-0.951}^{+0.979} ({-0.120})$ \\ 
$\tilde{\alpha}_4$ & $\mathcal{N}(0, 1)$ & $-0.347_{-0.938}^{+0.910} ({-1.281})$ & $-0.271_{-0.942}^{+0.913} ({-0.461})$ \\ 
$\tilde{\beta}_4$ & $\mathcal{N}(0, 1)$ & $-0.005_{-0.961}^{+1.027} ({0.271})$ & $-0.091_{-1.026}^{+0.983} ({-0.014})$ \\  
\hline
\end{tabular}
\end{table}}

Table \ref{tab:post} reports the values inferred in real space for the MCMC analysis with the priors adopted for the $[0.3,3]\hMpc$ and $[0.3,3]\hMpc$ scale cuts. $\mathcal{U}(a,b)$ represents a flat (uniform) prior over bounds $[a,b]$, while $\mathcal{N}(\mu, \sigma)$ represents a Gaussian prior of mean $\mu$ and standard deviation $\sigma$. $S_8$ and $\sigma_8$ are computed quantities. We note that these priors are given by the spline realizations shown in Figure \ref{fig:all_tomo}. The clustering redshifts approach takes into account galaxy bias from both samples and magnification effects. No combination with other photometric inference is performed, such that the constraints obtained depend only on the inclusion of clustering redshifts.

\begin{figure}[ht!]
    \centering
    \includegraphics[width=0.95\textwidth]{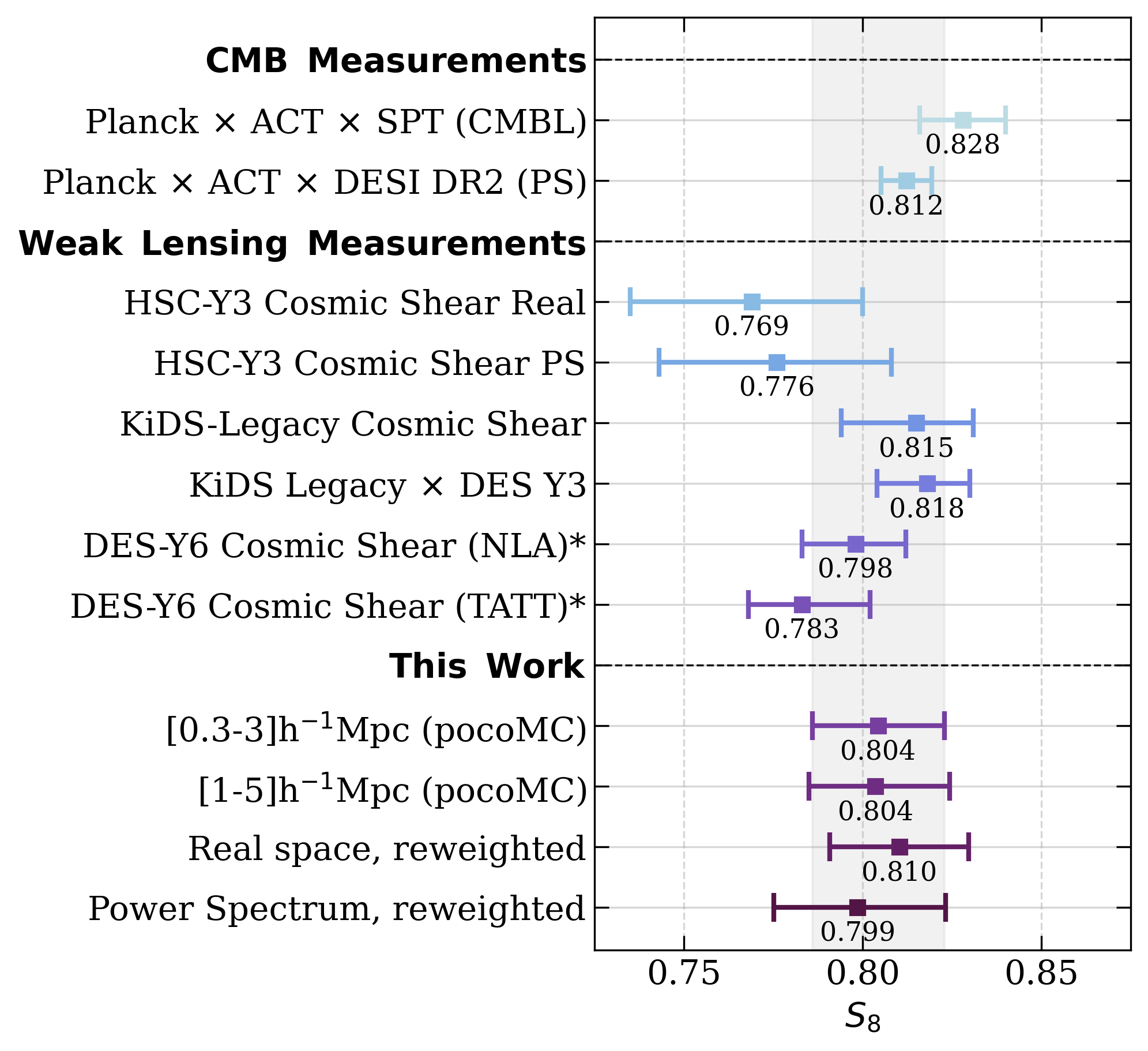}
    \caption{Measurements from different surveys measuring the growth of structure, from \textit{CMB Measurements} and \textit{Weak Lensing Measurements}. The results are further compared to the results obtained in this work, as well as combined with the $S_8$ constraints obtained. The  source of every cited measurement is discussed in text. We note that (*) measurements from DES-Y6 report the mean of the 1D posterior distribution as the fiducial measurement.}
    \label{fig:comparison_S8}
\end{figure}

The priors for all sampled parameters in the case of the fiducial $n(z)$ is reported in table \ref{tab:post}, alongside the posterior sampled values obtained for the other scale cuts. We note that the redshift $\Delta z_i$ priors depend on the chosen scale cut, as mentioned in appendix \ref{sec:appendix}.
Figure \ref{fig:comparison_S8} compares other recent $S_8$ constraints with respect to the constraints obtained in this work.  The most recent non-lensing CMB result is Planck $\times$ ACT $\times$ DESI DR2 constraints obtained from the power spectra measurements of ACT DR6 presented in \cite{Louis2025_ACTDR6_PS_S8}. The most recent CMB lensing result is Planck $\times$ ACT $\times$ SPT (CMBL): the CMB lensing constraints obtained from combining Planck, ACT and SPT lensing maps in work \cite{Qu2025_ACTSPTPLANCK_CMBL_S8}.
The figure shows measurements obtained with weak lensing from KiDS-Legacy \cite{Wright2025_KiDSLegacy}, DES Y6 \cite{DESy6_S8_CosmicShear}, HSC Y3 (in both real \cite{Li2023_HSCY3_CosmicShear} and Fourier \cite{Dalal2023CosmoShearHSC} space), and the combined measurement of KiDS-Legacy and DES Y3 presented in \cite{KiDS-Legacy_DESY3_Stolzner}, derived from the "KiDS-excised" DES datavector built in \cite{DES_KiDS}. Results from this paper inferred from the $n(z)$ re-calibration performed in \cite{ChoppinDeJanvry2025a} are presented next. The two scale cuts with all bins are measured with \texttt{pocoMC} sampling, as discussed in Section~\ref{sec:method:mcmc}. Results from importance re-weighting of the chains are also included.

%% file: sections/conclusion.tex
\section{Conclusion}
\label{sec:conclusion}

In this work, we provide cosmological constraints from cosmic shear with DESI-calibrated HSC Year 3 $n(z)$ tomographic bin distributions. We find $S_8=0.804^{+0.019}_{-0.018}$ when using information from clustering redshifts in the re-calibration analysis. Leveraging clustering information from quasars and emission line galaxies (above $z\gtrsim0.8$), the calibrated priors of Bins 3 and Bin 4 are shown to be significantly less biased compared to the self-calibration shift posteriors, as displayed in Figure~\ref{fig:corner_resampling_sc} and Figure~\ref{fig:param_sigma_shifts}. While our 
derived photometric bias is not zero, our results
show that the original photometric calibrations of HSC (\textbf{PhotZ}) are not very biased. 

The result can be compared to previous HSC Y3 result of $S_8=0.769^{+0.031}_{-0.034}$. Our result gives a $1.8$ reduction of error due to the improved clustering redshift calibrations, with 
the central value shifting considerably higher towards Planck cosmology. 
Our reanalysis makes HSC competitive with other Stage 3 lensing surveys such as DES Y3 and Y6 \cite{Amon2022CosmoShearDES, Secco2022CosmoShearDES, DESy6_S8_CosmicShear} and KiDS \cite{Wright2025_KiDSLegacy}. We investigate preliminary combinations with other weak lensing and CMB probes in appendix \ref{sec:appendix:combined}.
In contrast to the $S_8$ $\sim2\sigma$ tension with CMB results originally found for HSC Y3 in \cite{Li2023_HSCY3_CosmicShear, Dalal2023CosmoShearHSC}, the findings of this work show no evidence of significant deviations from Planck $S_8$ measurements for HSC. 

The compatibility with Planck results from the measurements of this analysis as well as the joint constraints from the Stage III weak lensing surveys suggest that the suspected "$S_8$ tension" \cite{DiValentino_S8_tension} is primarily systematics driven and has now been largely eliminated from all of the WL analyses. We emphasize that we focus on WL only analyses here: the analyses combining WL with galaxy clustering have additional potential systematics that may complicate their interpretation, and are not included here. 

We emphasize that survey level weak lensing calibration usually includes secondary calibration from photometry (see methods presented in \cite{DES2025_ClusteringSOM, SOMPZDES2024, wright2025kidslegacyClusterZ, HSCClusteringRau2023}). The redundancy allows for both better characterization of systematics through largely independent methods as well as combined statistics, thus leading to stringent constraints. In HSC Y3, this corresponds to the combination of \textbf{PhotZ} and \textbf{WX} results \cite{HSCClusteringRau2023}. However, \textbf{PhotZ} lacks spectroscopic redshift calibrations above $z\gtrsim1.2$ (from the specXphot sample), and no clustering results above that redshift either. 
While no independent results currently exist to compare against the clustering redshift results for tomographic bins 3 and 4, direct calibration methods \cite{lange2025unifiedphotometricredshiftcalibration, abby2026_inprep} implemented with DESI redshifts will provide alternative calibrations to the $n(z)$ distributions of weak lensing datasets. These methods 
will be able to validate 
results of our analysis. 

Our results suggest that clustering calibration of WL photometric redshifts plays a crucial role in achieving the goals of WL surveys. We expect the same to apply to the upcoming weak lensing releases of Stage III surveys such as the Dark Energy Survey Year 6 and HSC Public Data Release 4, as well as the upcoming releases of Stage IV weak lensing surveys such as Euclid \cite{EuclidI2024overview} and the Vera C. Rubin Legacy Survey of Space and Time \cite{LSST2019survey}. Assuming photometric redshifts can be reliably calibrated with the spectroscopic surveys such as DESI, these will provide new percent-level constraints on the growth of structure of the universe, further validating the standard cosmological model.

The code used for this work, the chains, and all the data to replicate the figures of this paper is made publicly available at \href{https://github.com/JeanCHDJdev/desi-y3-hsc/}{\textcolor{blue}{github.com/jeanchdjdev/desi-y3-hsc}}. For the specific cosmic shear analysis, this work is compiled under the \texttt{cosmic\_shear} directory of the repository.


%% file: sections/acknowledgements.tex
\acknowledgments
\label{sec:acknowledgements}

\hspace{0.85cm}JCdJ acknowledges support from Fondation CentraleSupélec and Fondation Ailes de France. BD acknowledges support from the Ambrose Monell Foundation, the Corning Glass Works Foundation Fellowship Fund, and the Institute for Advanced Study. SGG acknowledges that this work was performed in part at the Aspen Center for Physics, which is supported by National Science Foundation grant PHY-2210452 and a grant from the Alfred P Sloan Foundation (G-2024-22395). 
US is supported by NSF CDSE grant number AST-2408026 and NASA TCAN grant number 80NSSC24K0101. TZ is supported by Schmidt Sciences.
 

The Hyper Suprime-Cam (HSC) collaboration includes the astronomical communities of Japan and Taiwan, and Princeton University. The HSC instrumentation and software were developed by the National Astronomical Observatory of Japan (NAOJ), the Kavli Institute for the Physics and Mathematics of the Universe (Kavli IPMU), the University of Tokyo, the High Energy Accelerator Research Organization (KEK), the Academia Sinica Institute for Astronomy and Astrophysics in Taiwan (ASIAA), and Princeton University. Funding was contributed by the FIRST program from the Japanese Cabinet Office, the Ministry of Education, Culture, Sports, Science and Technology (MEXT), the Japan Society for the Promotion of Science (JSPS), Japan Science and Technology Agency (JST), the Toray Science Foundation, NAOJ, Kavli IPMU, KEK, ASIAA, and Princeton University. This paper makes use of software developed for Vera C. Rubin Observatory. We thank the Rubin Observatory for making their code available as free software at \url{http://pipelines.lsst.io/}. This paper is based on data collected at the Subaru Telescope and retrieved from the HSC data archive system, which is operated by the Subaru Telescope and Astronomy Data Center (ADC) at NAOJ. Data analysis was in part carried out with the cooperation of Center for Computational Astrophysics (CfCA), NAOJ. We are honored and grateful for the opportunity of observing the Universe from Maunakea, which has the cultural, historical and natural significance in Hawaii. 

This analysis made use of the following software packages: \texttt{CosmoSIS} (\url{https://cosmosis.readthedocs.io/en/latest/}) \cite{CosmoSISZuntz2015}, \texttt{pocoMC} (\url{https://pocomc.readthedocs.io/en/latest/}) \cite{karamanis2022pocoMC1, karamanis2022pocoMC2}, \texttt{getDist}(\url{https://getdist.readthedocs.io/en/latest/})\cite{getDistLewis2025}, \texttt{ChainConsumer} (\url{https://github.com/Samreay/ChainConsumer}) \cite{ChainConsumer_Hinton2016}, \texttt{matplotlib} \cite{Hunter2007Matplotlib}, \texttt{numpy} \cite{harris2020numpy}.

%% file: sections/appendix.tex
\section{Complementary parameter information}\label{sec:appendix:params}\label{sec:appendix}
In this appendix, we provide additional details to the parameters sampled during the cosmological inference runs. The priors are the same as \cite{Li2023_HSCY3_CosmicShear}, besides for the $\Delta z$ shift parameters, where we adopt Gaussian priors derived from the companion calibration work led in \cite{ChoppinDeJanvry2025a}. The fixed parameters in the cosmology are also conserved and are displayed in table \ref{tab:fixed}. The rest of the 23 free cosmological parameters priors can be found in Table \ref{tab:post}.

{\renewcommand{\arraystretch}{1.1}
\begin{table}[ht]
\centering
\caption{\label{tab:fixed}Fixed cosmological parameters for inference}
\begin{tabular}{l | l }
\hline
\textbf{Fixed parameters} & --\\
\hline
$\sum m_\nu$ (eV) & $0.06$\\
$w$ & $-1$\\
$w_a$ & $0$\\
$\Omega_\mathrm{k}$ & $0$\\
$\tau$ & $0.0851$ \\
\end{tabular}
\end{table}}

In table \ref{tab:post}, we report the priors used for the fiducial scale cut ($0.3-3\hMpc$). The priors used for the lager, $1-5\hMpc$ scale cut are in order : 
\begin{itemize}
    \item $\Delta z_1\sim\mathcal{N}(0, 0.013)$
    \item $\Delta z_2\sim\mathcal{N}(0, 0.014)$
    \item $\Delta z_3\sim\mathcal{N}(0, 0.028)$
    \item $\Delta z_4\sim\mathcal{N}(0, 0.016)$
\end{itemize}

\section{Chain processing}\label{sec:appendix:chain_processing}

\begin{figure}[h]
    \centering
    \includegraphics[width=0.98\linewidth]{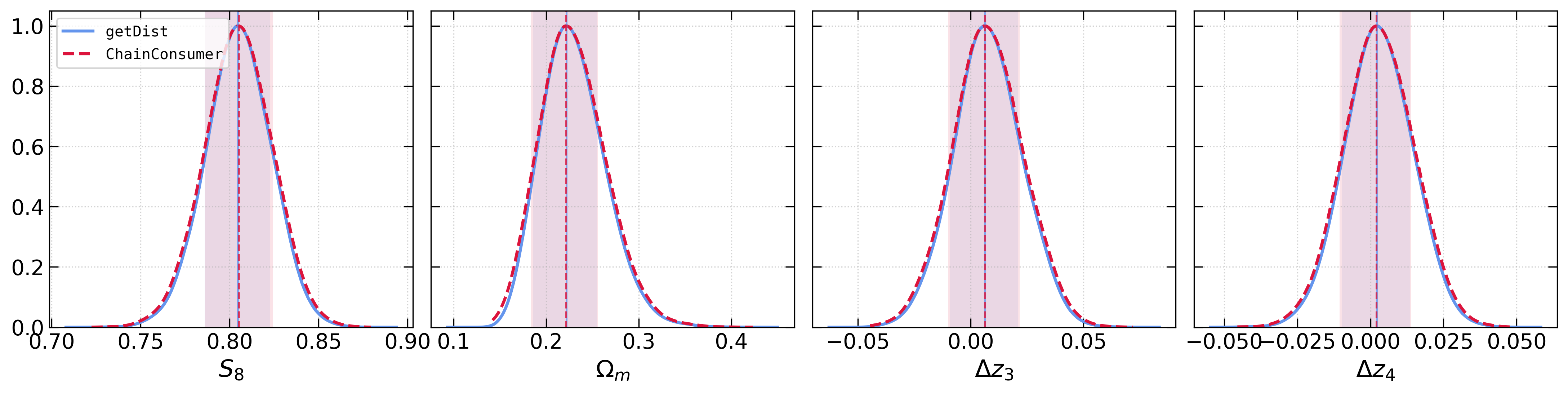}
    \caption{Comparison between posterior distributions for $S_8$, $\Omega_\mathrm{m}$, $\Delta z_3$, $\Delta z_4$ for $\texttt{getDist}$ and $\texttt{ChainConsumer}$. $\texttt{smooth\_scale\_1D}=-1$ (default parameter) is used for $\texttt{getDist}$. $\texttt{ChainConsumer}$ uses $\texttt{kde}=1$. The distributions shown come from the fiducial result using the $0.3-3\hMpc$ distributions.}
    \label{fig:appendix:getdist_cc_compare}
\end{figure}

MCMC chain processing is fiducially performed by $\texttt{getDist}$ with the default software parameters. The chains are also post-processed with $\texttt{ChainConsumer}$. We present posterior comparisons in Figure~\ref{fig:appendix:getdist_cc_compare}. We conclude there is no significant difference in processing, though we note that $\texttt{ChainConsumer}$ is less robust to edge effects if not corrected for, as mentioned in Fig. 23 of Li \textit{et al.} \cite{Li2023_HSCY3_CosmicShear}. This can be apparent for posterior distributions that are not well constrained. 

\section{Combined measurements}\label{sec:appendix:combined}

In this section, we present combined measurements from the different weak lensing Stage III surveys and CMB results. An important caveat to the following measurements is that they do not take into account any covariance between the surveys. This is especially important in the context of overlapping sky footprints, hence including multiple times the same galaxies. The following should therefore only be considered preliminary work to motivate thorough combination studies, e.g. as performed by \cite{DES_KiDS} when combining DES Y3 and KiDS-1000. 

\begin{figure}[h]
    \centering
    \includegraphics[width=0.8\linewidth]{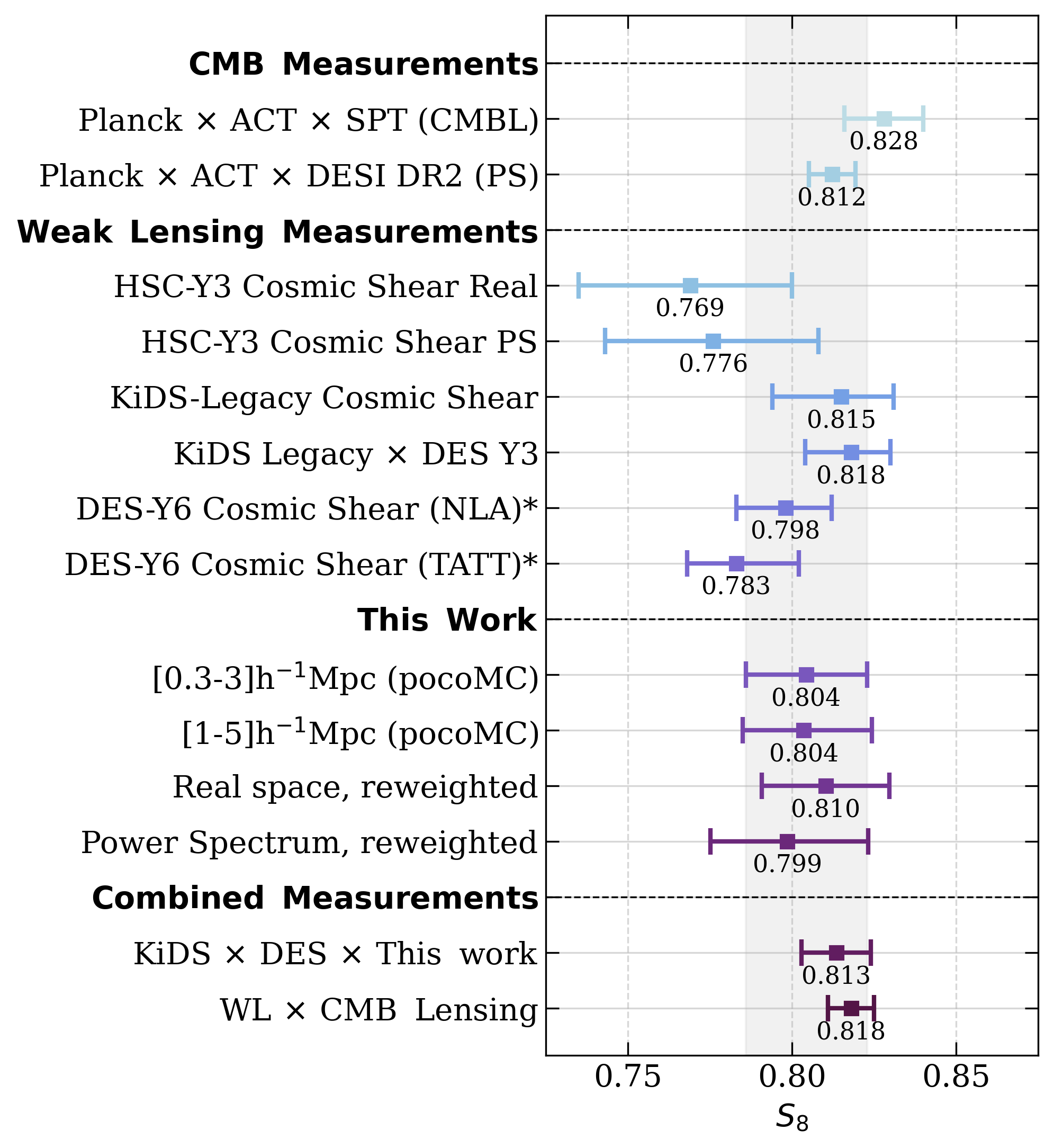}
    \caption{Visual comparison of reported $S_8$ values from Figure~\ref{fig:comparison_S8} alongside the combined measurements; first from DES Y3 $\times$ KiDS-Legacy $\times$ This Work, then combining that measurement with CMB Lensing.}
    \label{fig:appendix:s8_compare}
\end{figure}

The starting point for the combined measurements is the reported KiDS Legacy $\times$ DES Y3 \cite{KiDS-Legacy_DESY3_Stolzner} measurement, which uses the "KiDS-excised" DES Y3 data vector built in the DES Y3 $\times$ KiDS-1000 analysis \cite{DES_KiDS}. This is then naively combined with the fiducial measurement of this work ($S_8=0.805\pm{+0.018}\;(0.815)$), obtaining $S_8=0.813^{+0.009}_{-0.010}$ under the assumption that the surveys are independent. 
HSC Y3 has about 60\% area overlap with DES and KiDS, with effective density for HSC of $\sim15$ galaxies$/{\rm arcmin}^2$ \cite{Dalal2023CosmoShearHSC, Li2023_HSCY3_CosmicShear}. In contrast, DES Y3 and KiDS-Legacy respectively report $n_{\rm eff}\simeq5.59{\rm \;arcmin}^{-2}$ \cite{Secco2022CosmoShearDES} and $n_{\rm eff}\simeq8.79{\rm \;arcmin}^{-2}$ \cite{Wright2025_KiDSLegacy}. For DES Y6, this number becomes $n_{\rm eff}\simeq8.29{\rm \;arcmin}^{-2}$ \cite{DESy6_S8_CosmicShear}. This suggests around a third 
of galaxies in HSC already have a counterpart in DES or KiDS. One notes that HSC goes to fainter magnitudes, so the overlapping galaxies tend to be brighter and lay in the first redshift bin, which is less sensitive to the shear signal compared to the other higher redshift bins. This does not mean the higher tomographic bins are independent from DES+KiDS, although it is our assumption here for a naive combination of the WL surveys.


We further combine into the joint WL constraint with CMB Lensing information, obtaining $S_8=0.818\pm0.007$, a sub-percent constraint on $S_8$. This result can be compared to the most recent Planck $\times$ ACT $\times$ DESI DR2 ($S_8=0.812\pm0.007$), finding a similar value in both the mean and variance: we find no significant discrepancy between the two in the mean, and the combined constraining power of WL surveys is now comparable to that of the CMB. A comparison of the combined values as well as the fiducial measurements for this work is presented in Figure~\ref{fig:appendix:s8_compare}. 